\documentclass[prc,preprint]{revtex4}
\usepackage{graphics}

\newcommand{\be}{\begin{equation}}
\newcommand{\ee}{\end{equation}}
\begin{document}
\title{{\bf
Temperature dependence of magnetic susceptibility of nuclear
matter: lowest order constrained variational calculations}}
\author{{\bf M.
Bigdeli$^{1,3}$ \footnote{E-mail: m\underline\ \
bigdeli@znu.ac.ir}},  {\bf G.H. Bordbar $^{2,3}$\footnote{E-mail:
ghbordbar@shirazu.ac.ir}} and {\bf Z. Rezaei $^2$}}
\affiliation{ $^1$Department of Physics, Zanjan University, P.O.
Box 45195-313, Zanjan,
Iran\footnote{Permanent address}\\
$^2$Department of Physics, Shiraz University, Shiraz 71454,
Iran\footnote{Permanent address}\\
$^3$Research Institute for Astronomy and Astrophysics of Maragha,\\
P.O. Box 55134-441, Maragha, Iran}


\begin{abstract}
In this paper we study the magnetic susceptibility and other
thermodynamic properties of the polarized nuclear matter at finite
temperature using the lowest order constrained variational (LOCV)
method employing the $AV_{18}$ potential. Our results show a
monotonic behavior for the magnetic susceptibility which indicates
that the spontaneous transition to the ferromagnetic phase does
not occur for this system.
\end{abstract}
\maketitle


\section{INTRODUCTION}

The magnetic susceptibility is one of the most important magnetic
properties of the dense matter, its behavior specifies whether the
spontaneous phase transition to a ferromagnetic state occurs. This
transition in nuclear matter could have important consequences for
the physical origin of the magnetic field of the pulsars which are
believed to be rapidly rotating neutron stars with strong surface
magnetic fields in the range of $10^{12} -10^{13}$ Gauss
\cite{shap,paci,gold,navarro}. Considering different stages of the
neutron star formation at different temperatures, the study of the
magnetic properties of the polarized nuclear matter at finite
temperature is of special interest in the description of
protoneutron stars. A protoneutron star (newborn neutron star) is
born within a short time after the supernovae collapse. In this
stage the interior temperature of the neutron star matter is of
the order 20-50 MeV \cite{burro}.
The magnetic susceptibility of nucleonic matter is also an useful
quantity to estimate the mean free path of the neutrino in the
dense nucleonic matter which is relevant information for the
understanding of the mechanism underlying the supernova explosion
and the cooling process of the neutron stars \cite{iwamoto}.

There exists several possibilities of the generation of the
magnetic field in a neutron star. From the nuclear physics point
of view, such a possibility  has been studied by several authors
using different theoretical approaches [7-30], but the results are
still contradictory. In most calculations, the neutron star matter
is approximated by pure neutron matter at zero temperature. The
properties of the polarized neutron matter both at finite and zero
temperature have been studied by several authors
\cite{apv,dapv,bprrv}. Some calculations show that the neutron
matter becomes ferromagnetic for some densities
\cite{brown,rice,pandh,marcos,apv}. Some others,
 using the modern two-body and three-body realistic
interactions, show no indication of the ferromagnetic transition
at any density for the neutron matter and the asymmetrical nuclear
matter \cite{kutsb,fanto,vida,bprrv}. The results of another
calculation show both behaviors; the D1P force exhibits a
ferromagnetic transition whereas no sign of such transition is
found for D1 at any density and temperature \cite{dapv}.
The influence of the finite temperature on the antiferromagnetic
(AFM) spin ordering in the symmetric nuclear matter with the
effective Gogny interaction within the framework of a Fermi liquid
formalism has been studied by Isayev \cite{isay1,isay2}. We note
that in the symmetric nuclear matter corresponding to AFM spin
ordering, we have $\Delta\rho_{\uparrow\downarrow}=
(\rho_{n\uparrow}+\rho_{p\downarrow})-(\rho_{n\downarrow}+\rho_{p\uparrow})\neq0$
and
$\Delta\rho_{\uparrow\uparrow}=\rho_{\uparrow}-\rho_{\downarrow}=0$.
In this article, we use the lowest order constrained variational
(LOCV) formalism to investigate the possibility of the transition
to a ferromagnetic phase for the polarized hot symmetrical nuclear
matter.

The LOCV method has been developed to study the bulk properties of
the quantal fluids \cite{OBI1,OBI2,OBI3}. This technique has been
used for studying the ground state properties of finite nuclei and
treatment of isobars \cite{BHIM,MI1,MI2}.
 Modarres has extended the LOCV method to the finite temperature
 calculations and has applied it to the neutron matter,
 nuclear matter and asymmetrical nuclear matter in order to calculate the
 different thermodynamic properties of these systems \cite{Mod93,Mod95,Mod97,MM98}.
Few years ago, we calculated the properties of
 nuclear matter at zero and finite temperature using
the LOCV method with the new nucleon-nucleon potentials
\cite{BM97,BM98,MB98}.
The LOCV method has several advantages with respect to the other
many-body formalism. These are as follows: (i) Since the method is
fully self-consistent, it does not introduce any free parameters
into the calculations. (ii) It considers the constraint in the
form of a normalization constraint \cite{Feen} to keep the
higher-order terms as small as possible
\cite{OBI3,MI1,MI2,MM98,BM97} and it also assumes a particular
form for the long-range behavior of the correlation function in
order to perform an exact functional minimization of the two-body
energy with respect to the short-range behavior of the correlation
function. (iii) The functional minimization procedure represents
an enormous computational simplification over the unconstrained
methods (i.e. to parameterize the short-range behavior of the
correlation functions) which attempt to go beyond the lowest
order.

Recently, we have computed the properties of the polarized neutron
matter \cite{bordbig}, polarized symmetrical \cite{bordbig2} and
asymmetrical nuclear matters \cite{bordbig3} and also polarized
neutron star matter \cite{bordbig3} at zero temperature using the
microscopic calculations employing the LOCV method with the
realistic nucleon-nucleon potentials. We have concluded that the
spontaneous phase transition to a ferromagnetic state in these
matters does not occur.
We have also calculated the thermodynamic properties of the
polarized neutron matter at finite temperature \cite{bordbig4}
such as the total energy, magnetic susceptibility, entropy and
pressure using the LOCV method employing the $AV_{18}$ potential
\cite{wiring}.
 Our calculations do not show any transition to a ferromagnetic
 phase for a hot neutron matter.

In the present work, we intend to apply the LOCV calculation for
the polarized symmetrical nuclear matter at finite temperature
using the $AV_{18}$ potential.


\section{Finite temperature calculations for polarized nuclear matter with the LOCV method}

We consider a system of $A$ interacting nucleons with $A^{(+)}$
spin-up and $A^{(-)}$ spin-down nucleons. For this system, the
total number density ($\rho$) and spin asymmetry parameter
($\delta$) are defined as
\begin{eqnarray}
\rho&=&\rho^{(+)}+\rho^{(-)},\nonumber\\
\delta&=&\frac{\rho^{(+)}-\rho^{(-)}}{\rho}.
\end{eqnarray}
$\delta$ shows the spin ordering of the matter which can have a
value in the range of $\delta=0.0$ (unpolarized matter) to
$\delta=1.0$ (fully polarized matter).
To obtain the macroscopic properties of this system, we should
calculate the total free energy per nucleon, $F$,
\begin{eqnarray}\label{free}
       F=E-{\cal T}S^{(+)} -{\cal T}S^{(-)}.
 \end{eqnarray}
$E$ is total energy per nucleon and $S^{(i)}$ is the entropy per
nucleon corresponding to spin projection $i$,
 \begin{eqnarray}
 S^{(i)}(\rho,T)&=&-\frac{1}{A}\sum _{k}
 \{[1-n^{(i)}(k,{\cal T},\rho^{(i)})]\textrm{ln}[1-n^{(i)}(k,{\cal
 T},\rho^{(i)})]\nonumber\\&&
 +n^{(i)}(k,{\cal T},\rho^{(i)}) \textrm{ln} n^{(i)}(k,{\cal T},\rho^{(i)})\}.
\end{eqnarray}
where $n^{(i)}(k,{\cal T},\rho^{(i)})$ is the Fermi-Dirac
distribution function,
\begin{eqnarray}
n^{(i)}(k,{\cal T},\rho^{(i)})=\frac{1}{e^{\beta[\epsilon^{(i)}
(k,{\cal T},\rho^{(i)})-\mu^{(i)}({\cal T},\rho^{(i)})]}+1}\cdot
\end{eqnarray}
In the above equation $\beta=\frac{1}{k_B{\cal T}}$ , $\mu^{(i)}$
being the chemical potential which is determined at any adopted
value of the temperature $\cal T$, number density $\rho^{(i)}$ and
spin polarization $\delta$, by applying the following constraint,
 \begin{eqnarray}\label{chpt}
 \sum _{k}
 n^{(i)}(k,{\cal T},\rho^{(i)})=A^{(i)},
 \end{eqnarray}
and
 $\epsilon^{(i)}$ is the single particle energy of a nucleon. In
our formalism, the single particle energy of a nucleon with
momentum $k$ and spin projection $i$ is approximately written in
terms of the effective mass as follows \cite{apv,dapv,isay2}
 \begin{eqnarray}
 \epsilon^{(i)}(k,{\cal T},\rho^{(i)})=
 \frac{\hbar^{2}{k^2}}{2{m^{*}}^{(i)}(\rho,{\cal T})}+U^{(i)}({\cal
 T},\rho^{(i)}).
                \end{eqnarray}
In fact, we use a quadratic approximation for single particle
potential incorporated in the single particle energy as a momentum
independent effective mass. $U^{(i)}({\cal
 T},\rho^{(i)})$ is the
momentum independent single particle potential. We introduce the
effective masses, $m^{{*}{(i)}}$, as variational parameters
\cite{bordbig4,fp}. We minimize the free energy with respect to
the variations in the effective masses and then we obtain the
chemical potentials and the effective masses of the spin-up and
spin-down nucleons at the minimum point of the free energy. This
minimization is done numerically.

As it is also mentioned in the pervious section, for calculating
the total energy of the polarized symmetrical nuclear matter, we
use the LOCV method.
We adopt a trial many-body wave function of the form
\begin{eqnarray}
     \psi=\cal{F}\phi,
 \end{eqnarray}
where $\phi$ is the uncorrelated ground state wave function
(simply the Slater determinant of plane waves) of $A$ independent
nucleons and ${\cal F}={\cal F}(1\cdots A)$ is an appropriate
A-body correlation operator which can be replaced by a Jastrow
form i.e.,
\begin{eqnarray}
    {\cal F}={\cal S}\prod _{i>j}f(ij),
 \end{eqnarray}
in which ${\cal S}$ is a symmetrizing operator.
Now, we consider the cluster expansion of the energy functional up
to the two-body term \cite{clark},
 \begin{eqnarray}\label{tener}
           E([f])=\frac{1}{A}\frac{\langle\psi|H\psi\rangle}
           {\langle\psi|\psi\rangle}=E _{1}+E _{2}\cdot
 \end{eqnarray}
For the hot nuclear matter, the one-body term $E _{1}$ is
\begin{eqnarray}\label{ener1}
               E _{1}= E_{1}^{(+)}+E_{1}^{(-)},
 \end{eqnarray}
where
 \begin{eqnarray}
               E_1^{(i)}=\sum _{k}
               \frac{\hbar^{2}{k^2}}{2m}n^{(i)}(k,{\cal
               T},\rho^{(i)}).
                \end{eqnarray}
The two-body energy $E_{2}$ is
\begin{eqnarray}
    E_{2}&=&\frac{1}{2A}\sum_{ij} \langle ij\left| \nu(12)\right|
    ij-ji\rangle,
\end{eqnarray}
 where
\begin{eqnarray}
 \nu(12)=-\frac{\hbar^{2}}{2m}[f(12),[\nabla
_{12}^{2},f(12)]]+f(12)V(12)f(12).
\end{eqnarray}
In above equation, $f(12)$ and $V(12)$ are the two-body
correlation and potential.
In our calculations, we use the $AV_{18}$ two-body potential which
has the following form \cite{wiring},
\begin{equation}
V(12)=\sum^{18}_{p=1}V^{(p)}(r_{12})O^{(p)}_{12}, \label{v18}
\end{equation}
where
\begin{eqnarray}
O_{12}^{(p=1-18)}&=&1,\ {\bf\sigma_1}\cdot{\bf\sigma_2},\
{\bf\tau_1}\cdot{\bf\tau_2},\ ({\bf\sigma_1}\cdot{\bf\sigma_2})\
({\bf\tau_1}\cdot{\bf\tau_2}),\ S_{12},\
S_{12}({\bf\tau_1}\cdot{\bf\tau_2}),
\nonumber\\
&&{\bf L}\cdot{\bf S},\ {\bf L}\cdot {\bf
S}({\bf\tau_1}\cdot{\bf\tau_2}),\ {\bf L}^2,\ {\bf
L}^2({\bf\sigma_1}\cdot{\bf\sigma_2}),\
{\bf L}^2({\bf\tau_1}\cdot{\bf\tau_2}),\nonumber\\
&&{\bf
L}^2({\bf\sigma_1}\cdot{\bf\sigma_2})({\bf\tau_1}\cdot{\bf\tau_2}),\
({\bf L}\cdot {\bf S})^2,\ ({\bf L}\cdot {\bf
S})^2({\bf\tau_1}\cdot{\bf\tau_2}),\nonumber\\
&&{\bf T_{12}},\ ({\bf\sigma_1}\cdot{\bf\sigma_2}){\bf T_{12}},\
S_{12}{\bf T_{12}},\ (\bf\tau_{z1}+\bf\tau_{z2}). \label{operat}
\end{eqnarray}
In above equation, $S_{12}=[3({\bf\sigma_1}\cdot\hat
r)({\bf\sigma_2}\cdot\hat r)-{\bf\sigma_1}\cdot{\bf\sigma_2}]$ is
the tensor operator and ${\bf T_{12}}=[3({\bf\tau_1}\cdot\hat
r)({\bf\tau_2}\cdot\hat r) -{\bf\tau_1}\cdot{\bf\tau_2}]$ is the
isotensor operator. The above 18 components of the $AV_{18}$
two-body potential are denoted by the labels $c,\sigma,\tau,
\sigma\tau,t,t\tau,ls,ls\tau,l2,l2\sigma,l2\tau,l2\sigma\tau,ls2,ls2\tau,
T,\sigma T,tT$ and $\tau z$, respectively \cite{wiring}.
In the LOCV formalism, the two-body correlation $f(12)$ is
considered as the following form \cite{OBI3},
\begin{eqnarray}
f(12)&=&\sum^3_{k=1}f^{(k)}(r_{12})P^{(k)}_{12},
\end{eqnarray}
where
\begin{eqnarray}
P_{12}^{(k=1-3)}&=&\left (\frac{1}{4} -
\frac{1}{4}O^{(2)}_{12}\right ), \ \left (\frac{1}{2} +
\frac{1}{6}O^{(2)}_{12} + \frac{1}{6}O^{(5)}_{12}\right
),\nonumber\\&&\left (\frac{1}{4} + \frac{1}{12}O^{(2)}_{12} -
\frac{1}{6}O^{(5)}_{12}\right ).
\end{eqnarray}
The operators $O^{(2)}_{12}$ and $O^{(5)}_{12}$ are given in Eq.
(\ref{operat}).
Using the above two-body correlation and potential, after doing
some algebra we find the following equation for the two-body
energy,
\begin{eqnarray}\label{ener2}
E_{2} &=& \frac{2}{\pi ^{4}\rho }\left( \frac{h^{2}}{2m}\right)
\sum_{JLTSS_{z}T_{z}}\frac{(2J+1)(2T+1)}{2(2S+1)}[1-(-1)^{L+S+T}]\left|
\left\langle \frac{1}{2}\sigma _{z1}\frac{1}{2}\sigma _{z2}\mid
SS_{z}\right\rangle \right| ^{2} \nonumber \\&& \times \int
dr\left\{\left[{f_{\alpha }^{(1)^{^{\prime }}}}^{2}{a_{\alpha
}^{(1)}}^{2}(k_{f}r)+\frac{2m}{h^{2}}\left(\{V_{c}-3V_{\sigma }
+(V_{\tau }-3V_{\sigma \tau })(4T-3)
\right.\right.\right.\nonumber
\\&&\left.\left.\left.\ \ \ \ \ \ \ \ \ +(V_{T}-3V_{\sigma \tau })(4T)\}{a_{\alpha
}^{(1)}}^{2}(k_{f}r)\ +[V_{l2}-3V_{l2\sigma }
\right.\right.\right.\nonumber
\\&&\left.\left.\left.\ \ \ \ \ \ \ \ \
+(V_{l2\tau}-3V_{l2\sigma \tau })(4T-3)]{c_{\alpha
}^{(1)}}^{2}(k_{f}r)\right)(f_{\alpha }^{(1)})^{2}\right]
 \right. \nonumber
\\&& \left.\ \ \ \ \ \ \ \ \ \ +\sum_{k=2,3}\left[ {f_{\alpha }^{(k)^{^{\prime }}}}^{2}{a_{\alpha
}^{(k)}}^{2}+\frac{2m}{h^{2}}\left(\left\{V_{c}+V_{\sigma
}+(-6k+14)V_{t}+-(k-1)V_{ls}\right.\right.\right.\right. \nonumber
\\&&\left.\left.\left.\left.\ \ \ \ \ \ \ \ \ \ \ \ \ \ \ \ \ +[V_{\tau } +V_{\sigma \tau
}+(-6k+14)V_{tz}-(k-1)V_{ls\tau
}](4T-3)\right.\right.\right.\right. \nonumber
\\&& \left.\left.\left.\left.\ \ \ \ \ \ \ \ \ \ \ \ \ \ \ \ \ +[V_{T}+V_{\sigma \tau
}+(-6k+14)V_{tT}] [T(6T_z^2-4)]+ 2V_{\tau z}T_z\right\}{a_{\alpha
}^{(k)}}^{2}(k_{f}r) \right.\right.\right.\nonumber
\\&&\left.\left.\left. \ \ \ \ \ \ \ \ \ \ \ \ \ \ \ \ \ +[V_{l2}+V_{l2\sigma } +(V_{l2\tau }+V_{l2\sigma \tau
})(4T-3)]{c_{\alpha
}^{(k)}}^{2}(k_{f}r)\right.\right.\right.\nonumber
\\&&\left.\left.\left.\ \ \ \ \ \ \ \ \ \ \ \ \ \ \ \ \ +[(V_{ls2}+V_{ls2\tau
})(4T-3)]{d_{\alpha }^{(k)}}^{2}(k_{f}r)\right) {f_{\alpha
}^{(k)}}^{2}\right] \right.\nonumber \\&&\left. \ \ \ \ \ \ \ \ \
\ \ \ \ \ \ \ \ \ +\frac{2m}{h^{2}}[[(V_{ls\tau}-2(V_{l2\sigma
\tau }+V_{l2\tau })-3V_{ls2\tau })(4T-3)]\right. \nonumber
\\&& \left.\ \ \ \ \ \ \ \ \ \ \ \ \ \ \ \ \ \ +V_{ls}-2(V_{l2}+V_{l2\sigma
})-3V_{ls2}]b_{\alpha }^{2}(k_{f}r)f_{\alpha }^{(2)}f_{\alpha
}^{(3)} \right. \nonumber
\\&& \left.\ \ \ \ \ \ \ \ \ \ \ \ \ \ \ \ \ \
+\frac{1}{r^{2}}(f_{\alpha }^{(2)} -f_{\alpha
}^{(3)})^{2}b_{\alpha }^{2}(k_{f}r)\right\},
 \end{eqnarray}
where $\alpha=\{J,L,S,S_z\}$ and the coefficient ${a_{\alpha
}^{(1)}}^{2}$, etc., are as follows,
\begin{eqnarray}
     {a_{\alpha }^{(1)}}^{2}(x)=x^{2}I_{L,S_{z}}(x),
 \end{eqnarray}
\begin{eqnarray}
     {a_{\alpha }^{(2)}}^{2}(x)=x^{2}[\beta I_{J-1,S_{z}}(x)+\gamma
     I_{J+1,S_{z}}(x)],
 \end{eqnarray}
\begin{eqnarray}
           {a_{\alpha }^{(3)}}^{2}(x)=x^{2}[\gamma I_{J-1,S_{z}}(x)+\beta
           I_{J+1,S_{z}}(x)],
      \end{eqnarray}
\begin{eqnarray}
     b_{\alpha }^{(2)}(x)=x^{2}[\beta _{23}I_{J-1,S_{z}}(x)-\beta
     _{23}I_{J+1,S_{z}}(x)],
 \end{eqnarray}
\begin{eqnarray}
         {c_{\alpha }^{(1)}}^{2}(x)=x^{2}\nu _{1}I_{L,S_{z}}(x),
      \end{eqnarray}
\begin{eqnarray}
        {c_{\alpha }^{(2)}}^{2}(x)=x^{2}[\eta _{2}I_{J-1,S_{z}}(x)+\nu
        _{2}I_{J+1,S_{z}}(x)],
 \end{eqnarray}
\begin{eqnarray}
       {c_{\alpha }^{(3)}}^{2}(x)=x^{2}[\eta _{3}I_{J-1,S_{z}}(x)+\nu
       _{3}I_{J+1,S_{z}}(x)],
 \end{eqnarray}
\begin{eqnarray}
     {d_{\alpha }^{(2)}}^{2}(x)=x^{2}[\xi _{2}I_{J-1,S_{z}}(x)+\lambda
     _{2}I_{J+1,S_{z}}(x)],
 \end{eqnarray}
\begin{eqnarray}
     {d_{\alpha }^{(3)}}^{2}(x)=x^{2}[\xi _{3}I_{J-1,S_{z}}(x)+\lambda
     _{3}I_{J+1,S_{z}}(x)].
 \end{eqnarray}
In above equations, we have
\begin{eqnarray}
          \beta =\frac{J+1}{2J+1},\ \  \gamma =\frac{J}{2J+1}, \ \   \beta
          _{23}=\frac{2J(J+1)}{2J+1},
 \end{eqnarray}
\begin{eqnarray}
       \nu _{1}=L(L+1),\ \  \nu _{2}=\frac{J^{2}(J+1)}{2J+1},\ \  \nu
       _{3}=\frac{J^{3}+2J^{2}+3J+2}{2J+1},
      \end{eqnarray}
\begin{eqnarray}
     \eta _{2}=\frac{J(J^{2}+2J+1)}{2J+1}, \ \ \eta
     _{3}=\frac{J(J^{2}+J+2)}{2J+1},
 \end{eqnarray}
\begin{eqnarray}
     \xi _{2}=\frac{J^{3}+2J^{2}+2J+1}{2J+1},\ \   \xi
     _{3}=\frac{J(J^{2}+J+4)}{2J+1},
 \end{eqnarray}
\begin{eqnarray}
     \lambda _{2}=\frac{J(J^{2}+J+1)}{2J+1},\ \  \lambda
     _{3}=\frac{J^{3}+2J^{2}+5J+4}{2J+1},
 \end{eqnarray}
and
\begin{eqnarray}
       I_{J,S_{z}}(r,\rho,T)=\frac{1}{2\pi^{6}\rho^2}\int k_1^{2}dk_{1}k_2^{2}dk_{2}n_{i}
       (k_{1},T,\rho_{i})n_{j}(k_{2},T,\rho_{j})J_{J}^{2}(|k_{2}-k_{1}|r),
 \end{eqnarray}
where $J_{J}(x)$ is the Bessel's function .

 Now, we  minimize the two-body energy
Eq. (\ref{ener2}) with respect to the variations in the
correlation functions ${f_{\alpha}}^{(k)}$, but subject to the
normalization constraint \cite{OBI3,BM98},
\begin{eqnarray}\label{norm}
        \frac{1}{A}\sum_{ij}\langle ij\left| h_{S_{z}}^{2}
        -f^{2}(12)\right| ij\rangle _{a}=0\cdot
 \end{eqnarray}
In the case of polarized symmetrical nuclear matter, the Pauli
function $h_{S_{z}}(r)$ is as follows

\begin{eqnarray}
h_{S_{z}}(r)=
\left\{%
\begin{array}{ll}
\left[ 1-\frac{1}{2}\left( \frac{\gamma^{(i)}(r)
       }{\rho}\right) ^{2}\right] ^{-1/2} & ;\ \hbox{$S_{z}=\pm 1$} \\
    1 & ;\ \hbox{$S_{z}= 0$}
\end{array}%
\right.
\end{eqnarray}
 where
\begin{eqnarray}
\gamma^{(i)}(r)=\frac{1}{\pi^{2}}\int n^{(i)}(k,{\cal
T},\rho^{(i)})J_{0}(kr)k^2dk .
 \end{eqnarray}
 From
the minimization of the two-body cluster energy we get a set of
coupled and uncoupled Euler-Lagrange differential equations. The
Euler-Lagrange equations for uncoupled states are
\begin{eqnarray}
&g_\alpha^{(1)^{\prime\prime}}-\{\frac{a_\alpha^{(1)^{\prime\prime}}}
{a_\alpha^{(1)}}+\frac{m}{\hbar^2}
[V_c-3V_\sigma+(V_\tau-3V_{\sigma\tau})(4T-3)\nonumber\\&
+(V_T-3V_{\sigma T})[T(6T_z^2-4)]+2V_{\tau z}T_z+\lambda
]\nonumber\\& +\frac{m}{\hbar^2}(V_{l2}-3V_{{l2}\sigma}+
(V_{{l2}\tau}-3V_{{l2}\sigma\tau})(4T-3))
\frac{c_\alpha^{(1)^2}}{a_\alpha^{(1)^2}}\}g_\alpha^{(1)}=0,
\end{eqnarray}
while for the coupled states, these equations are written as
follows,
\begin{eqnarray}
&g_\alpha^{(2)^{\prime\prime}}-\{\frac{a_\alpha^{(2)^{\prime\prime}}}
{a_\alpha^{(2)}}+\frac{m}{\hbar^2} [V_c+V_\sigma+2V_t-V_{{ls}}
+(V_\tau+V_{\sigma\tau}+2V_{t\tau}
-V_{{ls}\tau})(4T-3)\nonumber\\& +(V_T+V_{\sigma
T}+2V_{tT})[T(6T_z^2-4)]+2V_{\tau z}T_z+\lambda ]\nonumber\\&
+\frac{m}{\hbar^2}[V_{l2}+V_{{l2}\sigma}
+(V_{{l2}\tau}+V_{{l2}\sigma\tau})(4T-3)]
\frac{c_\alpha^{(2)^2}}{a_\alpha^{(2)^2}}\nonumber\\&
+\frac{m}{\hbar^2}[V_{{ls}2}+
V_{{ls}2\tau}(4T-3)]\frac{d_\alpha^{(2)^2}}{a_\alpha^{(2)^2}}
+\frac{b_\alpha^2}{r^2a_\alpha^{(2)^2}}\}g_\alpha^{(2)}\nonumber\\&
+\{\frac{1}{r^2}-\frac{m}{2\hbar^2}[V_{ls}-
2V_{l2}-2V_{{l2}\sigma}-3V_{{ls}2}\nonumber\\&
+(V_{{ls}\tau}-2V_{{l2}\tau}-2V_{{l2}\sigma\tau}-
3V_{{ls}2\tau})(4T-3)]\}
\frac{b_\alpha^2}{a_\alpha^{(2)}a_\alpha^{(3)}}g_\alpha^{(3)}=0,
\end{eqnarray}
\begin{eqnarray}
&g_\alpha^{(3)^{\prime\prime}}-\{\frac{a_\alpha^{(3)^{\prime\prime}}}
{a_\alpha^{(3)}}+\frac{m}{\hbar^2} [V_c+V_\sigma-4V_t-2V_{ls}
+(V_\tau+V_{\sigma\tau}-4V_{t\tau}-2V_{{ls}\tau})(4T-3)\nonumber\\&
+(V_T+V_{\sigma T}-4V_{tT})[T(6T_z^2-4)]+2V_{\tau z}T_z+\lambda
]\nonumber\\&
+\frac{m}{\hbar^2}[V_{l2}+V_{{l2}\sigma}+(V_{{l2}\tau}
+V_{{l2}\sigma\tau})(4T-3)]
\frac{c_\alpha^{(3)^2}}{a_\alpha^{(3)^2}}\nonumber\\&
+\frac{m}{\hbar^2}[V_{{ls}2}+V_{{ls}2\tau}(4T-3)]\frac
{d_\alpha^{(3)^2}}{a_\alpha^{(3)^2}}
+\frac{b_\alpha^2}{r^2a_\alpha^{(2)^2}}\}g_\alpha^{(3)}\nonumber\\&
+\{\frac{1}{r^2}-\frac{m}{2\hbar^2}[V_{ls}-2V_{l2}-
2V_{{l2}\sigma}-3V_{{ls}2}\nonumber\\&
+(V_{{ls}\tau}-2V_{{l2}\tau}-2V_{{l2}\sigma\tau}
-3V_{{ls}2\tau})(4T-3)]\}
\frac{b_\alpha^2}{a_\alpha^{(2)}a_\alpha^{(3)}}g_\alpha^{(2)}=0,
\end{eqnarray}
where
\begin{eqnarray}
g_\alpha^{(i)}(r)=f_\alpha^{(i)}(r)a_\alpha^{(i)}(r).
\end{eqnarray}
The primes in the above equations mean differentiation with
respect to $r$.
 The Lagrange multiplier $\lambda $ is introduced by the
 normalization constraint, Eq. (\ref{norm}).
 Now, we can calculate the correlation functions by numerically
solving these differential equations and then using these
correlation functions, the two-body energy is obtained. Finally,
we can compute the energy and then the free energy of the system.

\section{Results and discussion }\label{NLmatchingFFtex}

\begin{figure}
\includegraphics{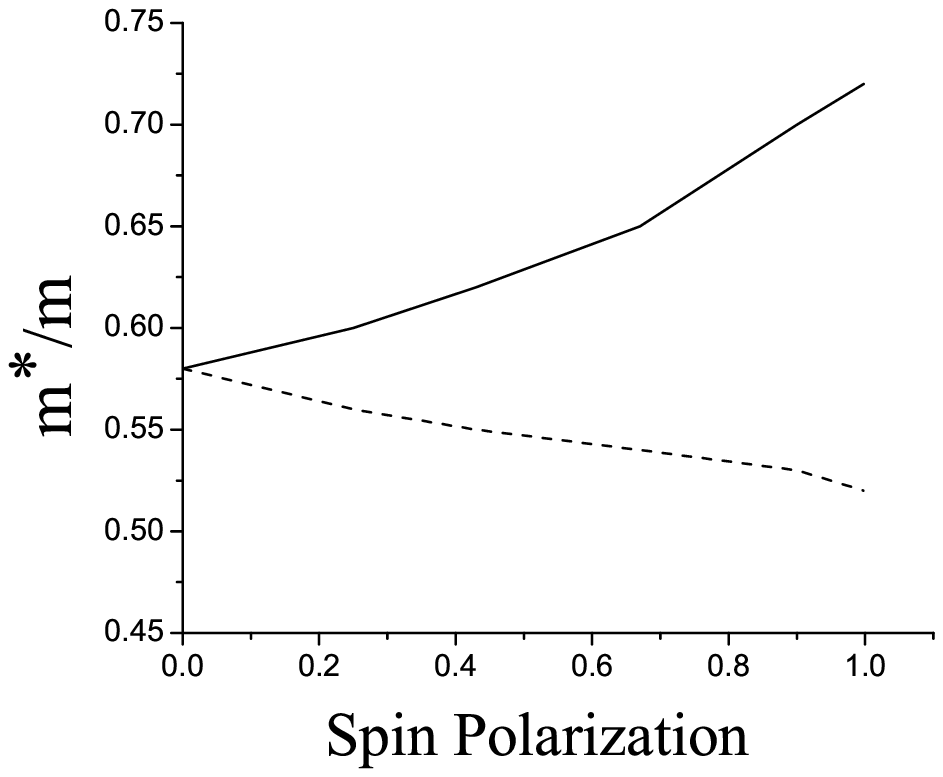}
 \caption{The effective mass of spin-up (full curve)
 and spin-down (dashed curve)
 nucleons versus the spin polarization ($\delta$) for
 density $\rho=0.5 fm^{-3}$ at ${\cal T}=20$ MeV.}
\label{effmass}
\end{figure}

We have presented the effective masses of the spin-up and
spin-down nucleons as functions of the spin polarization
($\delta$) at $\rho=0.5 fm^{-3}$ and  ${\cal T}=20$ MeV in Fig.
\ref{effmass}. It is seen that the difference between the
effective masses of spin-up and spin-down nucleons increases by
increasing the polarization. We also see that the effective mass
of spin-up nucleons increases by increasing the polarization
whereas the effective mass of spin-down nucleons decreases by
increasing the polarization.
These behaviors have been also seen for the effective mass of the
neutron in the case of spin polarized hot neutron matter
\cite{apv,dapv,bprrv,bordbig4}.
A similar qualitative behavior of the nucleon effective mass as a
function of the isospin asymmetric parameter
$\beta=\frac{\rho_n-\rho_p}{\rho_n+\rho_p}$ has been already found
in non-polarized isospin asymmetric nuclear matter, as it has been
already discussed in Refs. \cite{bomb1,li}.

\begin{figure}
\includegraphics{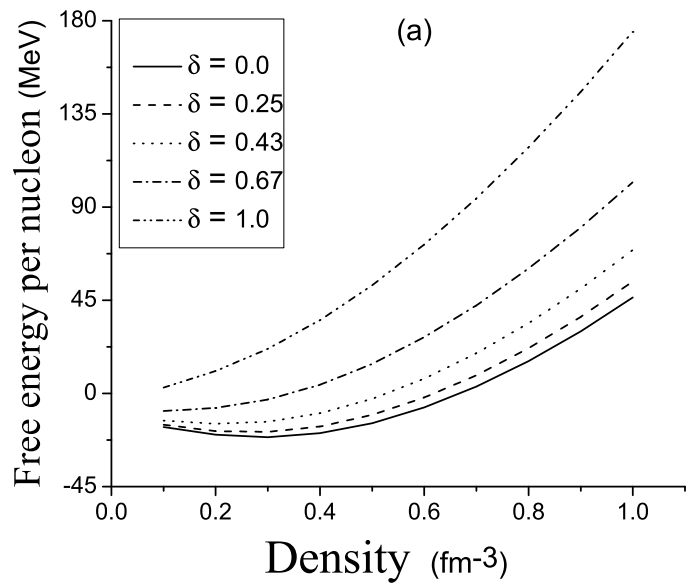}
\includegraphics{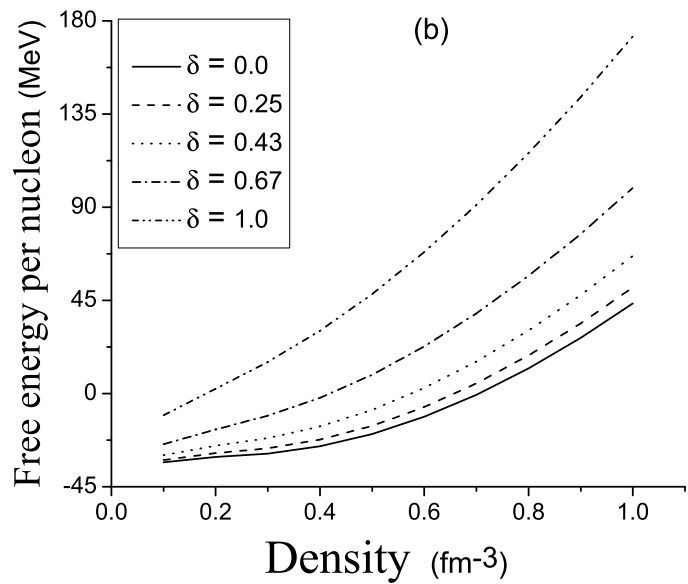}
 \caption{The free energy
per nucleon of the polarized hot nuclear matter as a function of
the total number density ($\rho$) for different values of the spin
polarization ($\delta$) at ${\cal T}=10$ (a) and ${\cal T}=20$ MeV
(b).} \label{freepol}
\end{figure}

\begin{figure}
\includegraphics{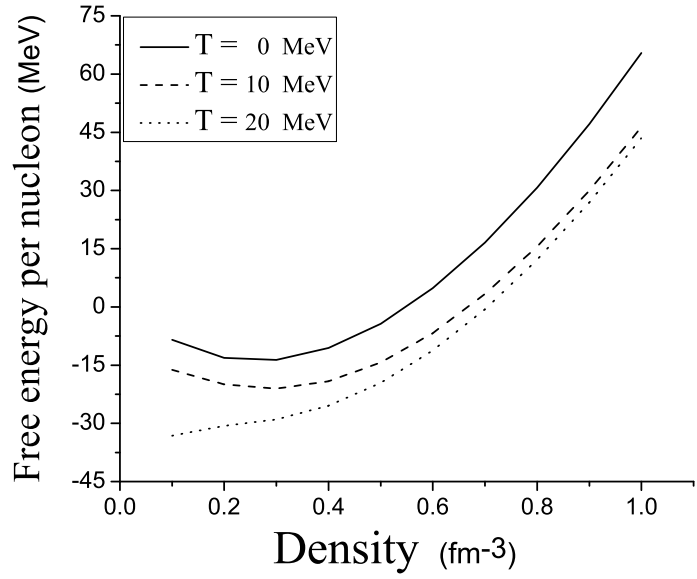}

 \caption{The free energy
per nucleon of the nuclear matter versus the total number density
($\rho$) for unpolarized case at ${\cal T}=0$, $10$ and $20$ MeV
.} \label{freeunpol}
\end{figure}

The free energy per nucleon of the polarized hot nuclear matter
versus the total number density ($\rho$) for different values of
the spin polarization ($\delta$) at ${\cal T}=10$ and $20$ MeV is
shown in Fig. \ref{freepol}. It can be seen that for each value of
the temperature, the free energy increases by increasing both
density and polarization. From Fig. \ref{freepol}, it is seen that
the free energy of the polarized hot nuclear matter decreases by
increasing the temperature.
 We have seen that above a certain values of the temperature and spin
polarization, the free energy does not show any bound states for
the polarized hot nuclear matter.
From Fig. \ref{freepol} we also see that for all given
temperatures there is no crossing of the free energy curves for
different polarizations and the difference between the free energy
of the nuclear matter at different polarizations increases by
increasing the density. This indicates that the spontaneous
transition to the ferromagnetic phase does not occur in the hot
nuclear matter.
We have also compared the free energy per nucleon of the
unpolarized case of the nuclear matter at different temperatures
in Fig. \ref{freeunpol}. It is seen that the free energy of
unpolarized nuclear matter decreases by increasing the
temperature. We can see that above a certain temperature, the free
energy does not show the bound state (the minimum point of the
free energy) for the unpolarized nuclear matter.

\begin{figure}
\includegraphics{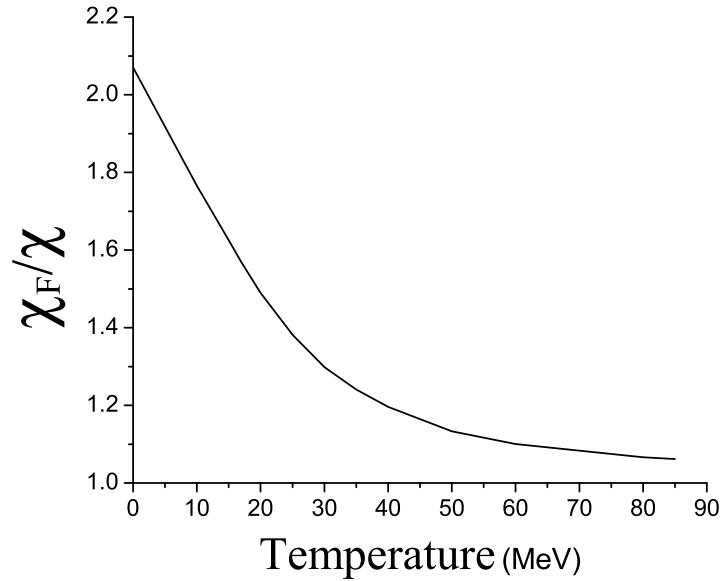}

 \caption{The magnetic
susceptibility of the hot nuclear matter versus the temperature at
$\rho=0.16 fm^{-3}$ .} \label{sustem}
\end{figure}

\begin{figure}
\includegraphics{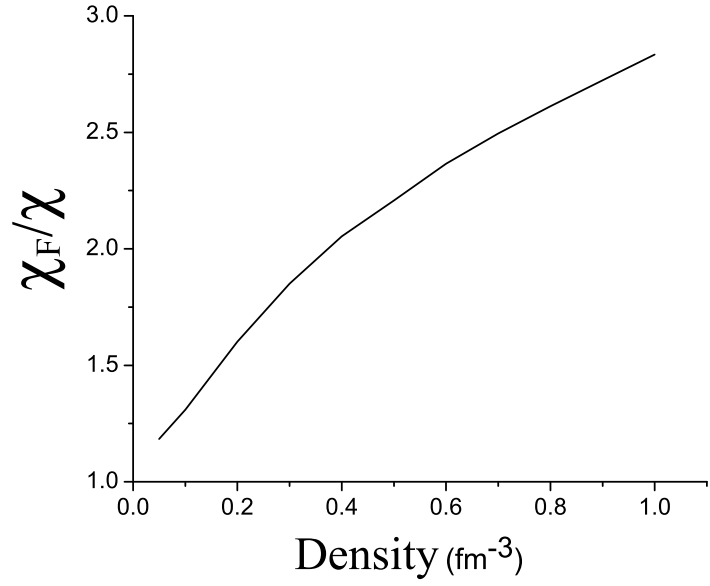}

 \caption{The magnetic
susceptibility of the hot nuclear matter versus the total number
density ($\rho$) at ${\cal T}=20$ MeV.} \label{susden}
\end{figure}

For the polarized hot nuclear matter, the magnetic susceptibility,
$\chi$, which characterizes the response of the system to the
magnetic field can be calculated by using the following relation,
\begin{eqnarray}\label{susep}
   \chi =\frac{\mu ^{2}\rho
}{\left( \frac{\partial^{2}F}{\partial \delta ^{2}}\right)
_{\delta =0}},
\end{eqnarray}
where $\mu$ is the magnetic moment of the nucleons.
  Fig. \ref{sustem} shows
the ratio ${{\chi_{F}}/{\chi}}$ as a function of the temperature
at $\rho=0.16 fm^{-3}$, where $\chi_{F}$ is the magnetic
susceptibility for a noninteracting Fermi gas. As it can be seen
from this figure, this ratio is inversely proportional to absolute
temperature without any anomalous change in its behavior. This
indicates that hot nuclear matter is paramagnetic.
The ratio ${{\chi_{F}}/{\chi}}$ has been also shown versus the
total number density at temperature  ${\cal T}=20$ MeV in Fig.
\ref{susden}. A magnetic instability would require
${{\chi_{F}}/{\chi} < 0}$. It is seen that the value of
${{\chi_{F}}/{\chi}}$ is always positive and monotonically
increasing up to highest density and does not show any spontaneous
phase transition to the ferromagnetic phase for the hot nuclear
matter.

\begin{figure}
\includegraphics{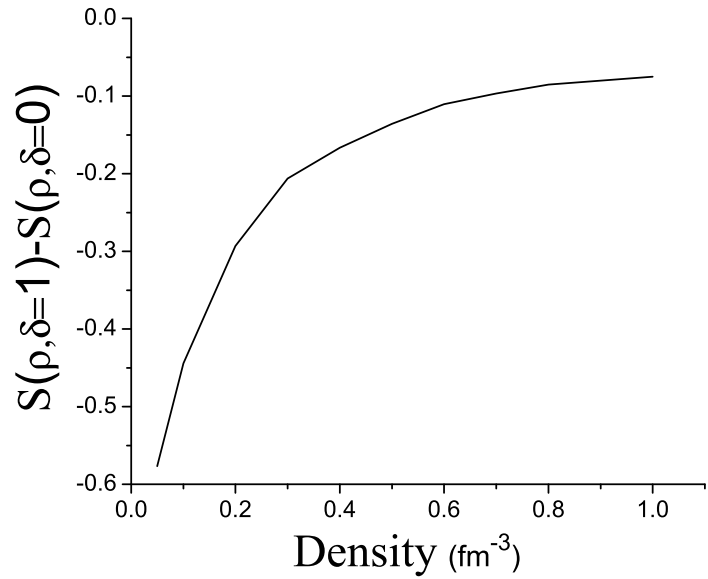}

 \caption{As Fig. 4 but for the entropy difference of the
fully polarized and the unpolarized cases.} \label{entden}
\end{figure}

\begin{figure}
\includegraphics{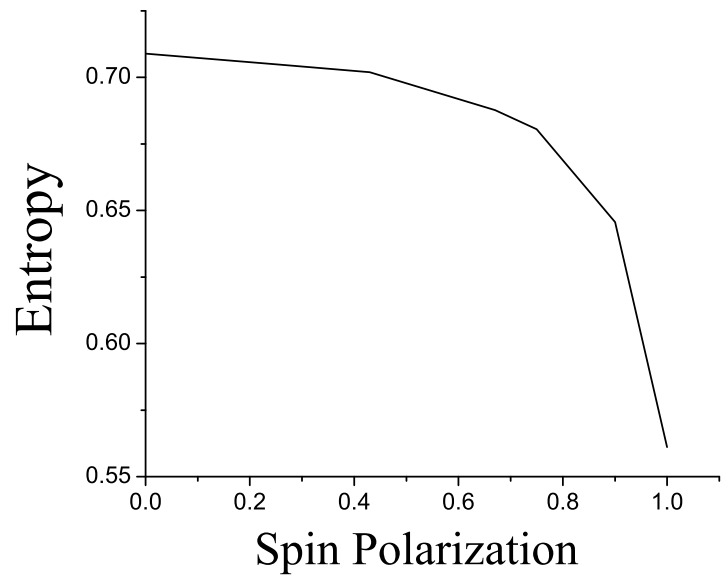}

 \caption{The entropy per nucleon
                as a function of the spin polarization ($\delta$)
                for density  $\rho=0.5 fm^{-3}$ at  ${\cal T}=20$ MeV .} \label{entpol}
\end{figure}

The difference between the entropy per nucleon of the fully
polarized and unpolarized cases of the nuclear matter is plotted
as a function of the total number density at ${\cal T}=20$ MeV in
Fig. \ref{entden}. It is seen that for all given values of the
density, this difference is negative. This shows that the fully
polarized case of hot nuclear matter is more ordered than the
unpolarized case.
 We also see that the magnitude of this deference decreases by
 increasing the density.
The entropy per nucleon of the polarized hot nuclear matter versus
the spin polarization for fixed density $\rho=0.5 fm^{-3}$ and
temperature ${\cal T}=20$ has been presented in Fig. \ref{entpol}.
It is shown that the entropy decreases by increasing the
polarization. It is also shown that the highest value of the
entropy occurs for the unpolarized case of the hot nuclear matter.
For the polarized hot nuclear matter, the following condition for
the effective mass prevents the anomalous behavior of the entropy
versus the spin polarization \cite{apv},
\begin{eqnarray}\label{emcon}
  \frac{m^{*}(\rho,\delta=1.0)}{m^{*}(\rho,\delta=0.0)}<2^{2/3},
\end{eqnarray}
where $m^{*}(\rho,\delta=1.0)$ and $m^{*}(\rho,\delta=0.0)$ are
the effective masses of the fully polarized and unpolarized
nuclear matter, respectively. This condition was first derived in
Ref. \cite{apv} for the particular case of the Skyrme interaction
where the effective mass is independent of the momentum and
temperature and therefore the single particle potential is purely
parabolic. In our approach, the effective mass depends on both
density and temperature but is independent of the momentum. In
other words, a similar rigorous condition can not be obtained
straightforwardly. However, within this approximation, one can use
the condition of Eq. (\ref{emcon}). From our result for the
effective mass at ${\cal T}=20 MeV$ for $\rho=0.5 fm^{-3}$ (Fig.
\ref{effmass}), we have found that this ratio is $1.24$. We see
that this value is smaller than the above limiting value which
indicates that the entropy of polarized case of the hot nuclear
matter is always smaller than the entropy of unpolarized case.
This so-considered "natural" behavior was also found in the case
of  Gogny \cite{dapv} and in the BHF analysis of Ref.
\cite{bprrv}. In contrast, for Skyrme forces the entropy per
particle of the polarized phase is seen to be higher than the
non-polarized one above a certain density \cite{dapv}.

\begin{figure}
\includegraphics{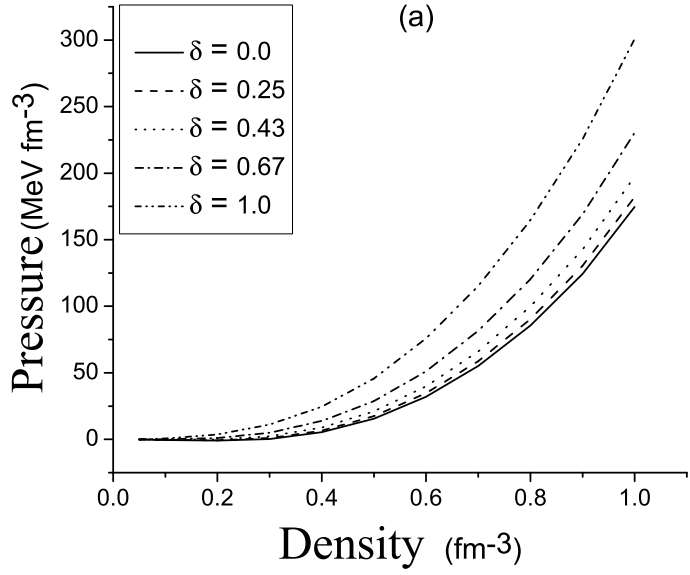}
\includegraphics{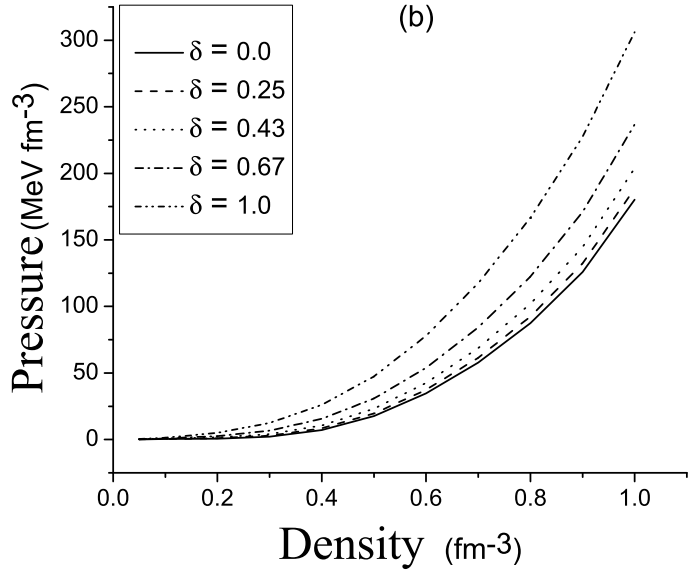}

\caption{The equation of state of the polarized hot nuclear matter
              for different values of the spin polarization ($\delta$)
              at ${\cal T}=10$ (a) and ${\cal T}=20$ MeV (b).}
\label{press}
\end{figure}

Finally, we have plotted the pressure of the polarized hot nuclear
matter as a function of the total number density ($\rho$) for
different polarizations at ${\cal T}=10$ and $20$ MeV in Fig.
\ref{press}. For all values of temperature and polarization, it is
seen that the pressure increases by increasing the density. For
this system, we see that at each temperature the equation of state
becomes stiffer as the polarization increases. For each
polarization, it is found that the pressure of the polarized hot
nuclear matter increases by increasing the temperature.
%
\section{Summary and Conclusions}
The lowest order constrained variational (LOCV) method has been
used for calculating the susceptibility of the polarized hot
nuclear matter and some of the thermodynamic properties of this
system such as the effective mass, free energy, entropy and the
equation of state. In our calculations, we have employed the
$AV_{18}$ potential. Our results show that the spontaneous
transition to the ferromagnetic phase does not occur for the hot
nuclear matter. We have seen that the spin polarization
substantially affects the thermodynamic properties of the hot
nuclear matter.


\acknowledgements {This work has been supported by Research
Institute for Astronomy and Astrophysics of Maragha. We wish to
thank Shiraz University and Zanjan University Research Councils.}



\newpage

\newpage


\newpage


\newpage


\newpage


\newpage


\newpage


\newpage


\end{document}